\documentclass{svmult}

\usepackage{graphicx} 
\usepackage{amsmath} 
\usepackage{amssymb} 
\usepackage{bbm} 

\renewcommand{\Re}{\operatorname{Re}}

\begin{document}

\title*{Is Bohmian Mechanics an empirically adequate theory?} 
\author{Kim Joris Bostr\"om} 
\institute{Kim Joris Bostr\"om \at University of M\"unster, Horstmarer Landweg 62b, \email{mail@kim-bostroem.de}}
\maketitle

\abstract{ Bohmian mechanics (BM) draws a picture of nature, which is completely different from that drawn by standard quantum mechanics (SQM): Particles are at any time at a definite position, and the universe evolves deterministically. Astonishingly, according to a proof by Bohm the empirical predictions of these two very different theories coincide. From the very beginning, BM has faced all kinds of criticism, most of which are either technical or philosophical. There is, however, a criticism first raised by Correggi et al. (2002) and recently strengthened by Kiukas and Werner (2010), which holds that, in spite of Bohm's proof, the predictions of BM do not agree with those of SQM in the case of local position measurements on entangled particles in a stationary state. Hence, given that SQM has been proven to be tremendously successful in the past, BM could most likely not be considered an empirically adequate theory. My aim is to resolve the conflict by showing that 1) it relies on hidden differences in the conceptual thinking, and that 2) the predictions of both theories approximately coincide if the process of measurement is adequately accounted for. My analysis makes no use of any sort of wavefunction collapse, refuting a widespread belief that an ``effective collapse'' is needed to reconcile BM with the predictions of SQM. }

\section{Introduction}

It is widely agreed that Bohmian mechanics (BM) makes the same predictions as standard quantum mechanics (SQM), which is the reason why both theories are usually considered as empirically equivalent ``interpretations'' of quantum mechanics. In his foundational papers, Bohm showed that the statistical predictions of BM for the outcome of measurements of arbitrary system properties coincide with those of SQM \cite{Bohm1952,Bohm1952a}. The empirical equivalence vitally clings on two things: 1) the validity and justification of the so-called ``quantum equilibrium hypothesis'' (QEH), which establishes a link between the wavefunction and a classical probability density on system configurations, and 2) the degree of precision of how the measurement process is accounted for. As for 1), it is still a matter of controversial debate whether the QEH has the status of a postulate or whether it can be independently derived from the underlying system dynamics \cite{Goldstein2009,Durr_et_al_1992,Durr_et_al_2013,Towler_et_al_2011,Valentini1991a,Valentini1991b}. In any case, Valentini showed that if the QEH is not exactly fulfilled, the predictions of BM may differ from those of SQM to a small but potentially measurable extent, which would make the empirical equivalence of BM and SQM experimentally testable \cite{Valentini1991a,Valentini1991b}. Here I shall not be concerned with the QEH, but I will rather investigate a recently revived controversy related to the secondly mentioned critical issue about measurement.

The consensus among most physicists concerned with BM seems to be that potentially deviating predictions of BM are ``harmless'' in that they can be made arbitrarily small and do not lead to inconsistencies or gross empirical disagreement with the standard theory. Proponents of BM take it as an advantage of their theory that measurement is not an additional concept fundamentally different from ordinary physical processes, as is the case in SQM, but that measurement is a specially designed but otherwise ordinary physical process that must be explicitly included in the description to obtain empirically valid results. As long as the predictions of BM can be made arbitrarily close to those of SQM by taking into account a suitably designed measurement interaction between the system of interest and a macroscopic measurement device, there would be no reason to doubt the empirical equivalence of both theories.

However, quite recently this consensus has been seriously put into question. Kiukas and Werner have published a thorough analysis of CHSH inequalities, showing (amongst other things) that BM leads to rather drastically different predictions for position measurements on stationary entangled states \cite{Kiukas_et_al_2010} . After publication the authors found that their line of reasoning had already been followed by Correggi and Morchio \cite{Correggi_et_al_2002}, but their new analysis provided theoretical insights valuable and novel enough to be published as a major contribution to the understanding of standard quantum mechanics and its relation to hidden-variable theories such as BM and Nelson's theory of stochastic evolution of classical particle configurations \cite{Nelson1966}. If the authors (Kiukas and Werner, as well as Correggi and Morchio) are right in what they say, then the Nelson-Bohm class of theories can no longer be considered as empirically equivalent to SQM, and moreover it would be relatively easy to devise an \emph{experimentum crucis} to decide which one of these theories is empirically valid. Kiukas and Werner leave no doubt which theory they think will pass the test (SQM), and certainly even the hardest followers of BM will acknowledge that in this case SQM can hardly be expected to fail. There is a recent defense by D\"urr et al., in which the authors hold that BM yields the same predictions as SQM even for the critical scenario under consideration, if only the ``collapse'' induced by the measurement interaction is correctly taken into account \cite{Durr_et_al_2014}. However, a similar reaction has already been anticipated by Kiukas and Werner, and in their paper they argue against the salutary role of the collapse in such a scenario. There has been no further contribution since, and although both sides presumably consider the issue as settled in favor of their view, the interested neutral public may largely be left with some confusion and the unsatisfactory feeling of witnessing a stalemate situation.

Here I aim to remove the confusion and settle the issue in a way that both sides may agree upon. In my view, the main reasons for the apparently conflicting conclusions drawn from the analysis of the critical scenario under consideration (in the following referred to as ``the scenario'') is found in 1) the abstract generality and the sophisticated non-mainstream methodology (C{*}-algebra) of the analysis carried out by Kiukas and Werner, 2) the rather sketchy defense by D\"urr et al., in which the ``collapse'' takes a central role in reconciling the seemingly different predictions, and 3) the differences inherent in the conceptual languages used by either side. I address the first point by re-analyzing the scenario in the framework of either theory, SQM and BM, while using only mainstream mathematics that every informed reader will be familiar with. Secondly, I provide detailed calculations in the framework of BM without making use of any sort of ``collapse''. As for the third point, I will seek to clarify to what extent the \emph{physical meaning} that each party in this conflict gives to their formal expressions inherently differs, so that it becomes more evident where exactly the protagonists may \emph{think} they talk about the same thing, although they're not. The result of my re-analysis is that BM is to the same extent empirically equivalent to SQM as it was before the conflict was brought about.

\section{Potential sources of confusion}

Let us start with some potential sources of confusion that may obscure the view on the conceptual differences between BM and SQM, and which may prevent a proper resolution of the apparent conflict.

\subsection{Position}

When Bohmians talk about the position of particles, they mean something else than adherents of SQM. The term ``position'' refers in BM to a classical variable, while in SQM it refers to an operator. As John Bell has put it, position in BM represents a \emph{beable} rather than an observable \cite{Bell1986}. The particle \emph{is} at any time at a specific position, whether it is being measured or not. These beables, the particle positions, are the ``hidden variables'' of Bohm's theory. Another beable, according to Bell, is the wavefunction. Whether being measured or not, the system \emph{has} a certain wavefunction. These beables are \emph{subject to}, and are generally \emph{affected by}, measurements, that is, their value is subject to uncontrollable and potentially drastic alteration during a measurement. In contrast to beables, operator-observables represent an \emph{experimental procedure} amounting to the measurement of a specific system property by an external macroscopic apparatus. This is what Bohmians mean when they say that observables are \emph{contextual}, and it is virtually the same idea that already had been brought up by Nils Bohr in support of his principle of \emph{complementarity}. In his analysis of the EPR paradox, Bohr writes \cite[pp 699-700]{Bohr1935}: \begin{quote} In fact to measure the position of one of the particles can mean nothing else than to establish a correlation between its behavior and some instrument rigidly fixed to the support which defines the space frame of reference. \end{quote} Beables, on the other hand, are non-contextual. So when a Bohmian utters the phrase ``the particle is at time $t$ at position $\boldsymbol{x}$'', he does not mean that the particle is being measured at position $\boldsymbol{x}$, nor does he mean that the system is in the eigenstate $|\boldsymbol{x}\rangle$ of the position operator. Rather, he is referring to a classical hidden variable $\boldsymbol{x}$ that is not attached to a quantum state or an operator. It is an \emph{additional concept} that simply does not exist in the terminology of SQM, and which can, however, be put into analogy with the ``true state'' of a classical system being described by a probability distribution over phase space. In classical statistical mechanics there is no doubt that the system is at any time in an exact state, which is represented by a point in phase space. Still, at the descriptive level the system state may well be represented by a probability distribution on phase space. These two descriptive elements of statistical mechanics, the probability distribution and the point in phase space, do not rival each other. Rather, the probability distribution simply captures the ignorance of the observer about the true state of the system. In practice, only a few macro-observables like volume, pressure, and temperature, are known to the observer, and these macro-observables fix a probability distribution over phase space that matches the constraints. The probability distribution and the point in phase space are also referred to as the \emph{macrostate} and the \emph{microstate}, respectively. It is helpful to recall the analogy of macrostate and microstate with wavefunction and system configuration, respectively, when trying to make sense of statements formulated within BM. The analogy is not perfect, though, because in BM the wavefunction is taken to \emph{objectively exist}, while in statistical mechanics the probability distribution is just a mathematical tool to calculate probabilities. In BM, the wavefunction guides the particles, while it is not asserted in statistical mechanics that the probability distribution in any way ``guides'' the particles. The conceptual stance of SQM is considerably different from that of BM. In SQM, the microstate may rather be put into analogy with the wavefunction, while the macrostate may then be put into analogy with the density matrix. In other words, in SQM the wavefunction already \emph{is} the complete description of the system, there is no further level of detail in the structure of reality, or at least any such further detail would be physically irrelevant and should not be part of a physical theory.

\subsection{Collapse}

In SQM, measurement is considered fundamentally different from ordinary Schr\"odinger evolution. It is a discontinuous process described by a projection and a subsequent renormalization, also referred to as the ``collapse of the wavefunction'', while the ordinary Schr\"odinger evolution is a continuous process described by a unitary transformation. The ontological status of, and the relation between, these two kinds of process is the issue of the infamous ``measurement problem''. Although hardly anybody seems to like the ``collapse'', it is there in the axioms of SQM, and it does a perfect job when it comes to statistical predictions of measurement outcomes.

On the other hand, BM is a collapse-free theory. Measurement is just a specially designed but otherwise ordinary physical process involving a short and strong interaction between the system of interest and a macroscopic measurement device. When Bohmians mention the ``collapse'' in their analysis of a given scenario within the framework of BM, then what they really mean is an ``effective'' collapse. The effective collapse refers to a mathematical simplification of the physical description. More precisely, it amounts to cutting away ``empty branches'' from the wavefunction, which are components of a suitably chosen decomposition of the wavefunction that do not guide any particles. Since they do not guide any particles, they have no effect on the outcome of future measurements, and so they can safely be removed from the description, although ontologically they are still there. However, it is always possible to leave the empty branches in the description and ignore the effective collapse altogether, without affecting the statistical predictions of measurement outcomes. Only, the calculations will become more complicated, because the empty branches have to be taken everywhere into account. In this sense, the effective collapse is a helpful, but not an essential, part of the Bohmian methodology. There is, however, a sense in which the effective collapse is more than just a mathematical convenience: It actually yields an \emph{explanation} for the seemingly random, discontinuous jump of the wavefunction introduced by a measurement process. We shall come back to this later on in the Discussion.

\subsection{CHSH inequalities}

Every hidden-variable theory faces the challenge of CHSH inequalities \cite{Clauser_et_al_1969}, which are generalizations of Bell's inequality \cite{Bell1964}. If a candidate theory does not violate a CHSH inequality under circumstances where standard quantum mechanics (SQM) does, then it is not empirically equivalent to SQM, and most probably, at least according to the experimental evidence collected over the past decades, it is not empirically valid. Maximal violations of CSHS inequalities involve correlations between the results of local measurements performed on entangled systems.

The first and most famous case of CHSH violation has been investigated by Bell \cite{Bell1964}, where he showed that no local hidden-variable theory can violate a special CHSH inequality, the Bell inequality, and is thus in conflict with SQM. Subsequent experiments \cite{Aspect_et_al_1982} have corroborated the predictions of SQM and thus have ruled out any local hidden variables theory as empirically false. Ironically, Bell himself was a great defender of Bohmian mechanics (BM), which \emph{is} a hidden-variable theory \cite{Bell1971,Bell1982,Bell1990}. As he and other defenders of BM have argued, the crucial feature that helps BM circumvent impossibility proofs, is that it is a \emph{nonlocal} theory. Specifically, the velocity of a given particle generally depends on the instantaneous position of remote particles. Moreover, and of equal importance, Bell pointed out that the operator-observables considered in standard quantum mechanics do not represent properties of the isolated system alone, but rather they depend on the entire measurement arrangement, a feature that has been termed \emph{contextuality.}

Since Bohmian mechanics has a joint probability distribution for the positions of particles at all times, the correlation between these positions at distinct times cannot violate a CHSH inequality and thus BM appears to be in conflict with the standard theory. The defenders of BM have argued that their theory nevertheless reproduces the same predictions as SQM once the measurement process is adequately accounted for \cite{Berndl_et_al_1996,Durr_et_al_2014}. Interestingly, this is basically the same route that also Bohr took in his attempt to defend SQM against the challenge posed by the EPR paradox \cite{Bohr1935}. So, it seems the same conflicting intuitions about the role of measurement and the nature of reality are raising their head against each other again and again.

\section{The critical scenario}

Let us give a brief, non-formal description of the critical scenario where the predictions of SQM and of BM are claimed to differ \cite{Correggi_et_al_2002,Kiukas_et_al_2010}. Alice and Bob, the usual suspects in modern quantum mechanical scenarios, are at remote sites and each one has a particle under their control. The particles are entangled with each other in a local energy eigenbasis, so that the state of the total system is itself in an eigenstate of the total energy. As the particles are now remote from each other, there is no interaction between them any more, or the interaction is so weak that it can be ignored, and hence no further entanglement between them is created. Because the total system is in an energy eigenstate, it is stationary. In Bohmian mechanics this implies that the probability density for the position of the particles does not change over time. Because the particles are energetically decoupled, the marginal densities of Alice's and Bob's particle remain constant in time, too, and there can be no time-dependent correlation between the position of both particles. In contrast, when calculating the expectation value of the product of the position observables at distinct times, which gives the two-time correlation function for the measured positions of the particles, one finds that there is a time-dependent correlation between the two measurements. Hence, either BM is empirically false (tacitly assuming that SQM does not fail), or the particles in BM ``are not where they are measured''. Either way, it seems that BM looses out against SQM.

\subsection{Analysis in standard quantum mechanics}

Alice and Bob each possess an identical copy of a system with a local Hamiltonian $\hat{H}$ and local energy eigenstates $|n\rangle$, so that \begin{equation} \hat{H}|n\rangle=E_{n}|n\rangle. \end{equation} The compound system of Alice and Bob has the total Hamiltonian $\hat{H}_{AB}=\hat{H}_{A}+\hat{H}_{B}$, where $\hat{H}_{A}=\hat{H}\otimes\mathbbm1$ and $\hat{H}_{B}=\mathbbm1\otimes\hat{H}$. Alice and Bob together prepare the system at the initial time $t=0$ in an entangled energy eigenstate \begin{equation} |\Psi_{0}\rangle=\frac{1}{\sqrt{2}}[|0\rangle|1\rangle+|1\rangle|0\rangle],\label{eq:initialstate} \end{equation} so that $\hat{H}_{AB}|\Psi_{0}\rangle=(E_{0}+E_{1})|\Psi_{0}\rangle$. Since $|\Psi_{0}\rangle$ is an energy eigenstate, it is stationary, so in the Schr\"odinger picture we have a time-dependent wavefunction $|\Psi_{t}\rangle=e^{-i(E_{0}+E_{1})t}|\Psi_{0}\rangle$ with an oscillating global phase that is irrelevant and can be ignored. Note that here and in the following we set $\hbar=1$ to simplify the calculations.

In the Heisenberg picture, the measurement of an observable $\hat{A}$ at time $t$ is represented by the operator $\hat{A}(t)=\hat{U}^{\dagger}(t)\hat{A}\hat{U}(t)$, so that $\langle\hat{A}\rangle(t)=\langle\Psi_{0}|\hat{A}(t)|\Psi_{0}\rangle$. We can then write the two-time correlation function between any two observables $\hat{A}$ and $\hat{B}$ as \begin{equation} \langle\hat{A}\hat{B}\rangle(t_{1},t_{2})=\langle\Psi_{0}|\hat{A}(t_{1})\hat{B}(t_{2})|\Psi_{0}\rangle. \end{equation} At time $t_{1}>0$ Alice measures the local observable $\hat{A}_{A}=\hat{A}\otimes\mathbbm1$ on her particle, so her measurement will obey the statistics of the operator $\hat{A}_{A}(t_{1})=e^{i\hat{H}t_{1}}\,\hat{A}\,e^{-i\hat{H}t_{1}}\otimes\mathbbm1$. At some later time $t_{2}>t_{1}$ Bob measures the local observable $\hat{B}_{B}=\mathbbm1\otimes\hat{B}$, and his measurement will obey the statistics of the operator $\hat{B}_{B}(t_{2})=\mathbbm1\otimes e^{i\hat{H}t_{2}}\,\hat{B}\,e^{-i\hat{H}t_{2}}$. Since Bob's measurement applies to a different subsystem, his and Alice's operators commute for all times $t_{1},t_{2}$, so that the product of both observables obeys the statistics of the common operator \begin{equation} \hat{A}_{A}(t_{1})\hat{B}_{B}(t_{2})=e^{i\hat{H}t_{1}}\,\hat{A}\,e^{-i\hat{H}t_{1}}\otimes e^{i\hat{H}t_{2}}\,\hat{B}\,e^{-i\hat{H}t_{2}}. \end{equation} The expectation value of this operator yields the two-time correlation function \begin{equation} \langle\hat{A}_{A}\hat{B}_{B}\rangle(t_{1},t_{2})=\langle\Psi_{0}|\hat{A}_{A}(t_{1})\hat{B}_{B}(t_{2})|\Psi_{0}\rangle, \end{equation} which yields for the initial state \eqref{eq:initialstate} \begin{equation} \begin{split}\langle\hat{A}_{A}\hat{B}_{B}\rangle(t_{1},t_{2}) & =\Re\left\{ \langle0|\hat{A}|1\rangle\langle1|\hat{B}|0\rangle e^{i\varDelta E(t_{2}-t_{1})}\right\} \\ & \phantom{=}+\frac{1}{2}\left[\langle0|\hat{A}|0\rangle\langle1|\hat{B}|1\rangle+\langle1|\hat{A}|1\rangle\langle0|\hat{B}|0\rangle\right]. \end{split} \label{eq:twotimeAB} \end{equation} So the two-time correlation function oscillates with the temporal distance between the two measurements, with a frequency proportional to the energy difference $\varDelta E=E_{1}-E_{0}$, and its value does not depend on the temporal ordering of the two measurements. For the particular case that Alice and Bob both measure the position of their particle, we have $\hat{A}=\hat{B}=\hat{\boldsymbol{x}}$, and the two-time correlation function of their measurements yields

\begin{equation} \begin{split}\langle\hat{\boldsymbol{x}}_{A}\hat{\boldsymbol{x}}_{B}\rangle(t_{1},t_{2}) & =|\langle0|\hat{\boldsymbol{x}}|1\rangle|^{2}\cdot\cos(\varDelta E(t_{2}-t_{1}))\\ & \phantom{=}+\langle0|\hat{\boldsymbol{x}}|0\rangle\langle1|\hat{\boldsymbol{x}}|1\rangle, \end{split} \label{eq:twotimeQM} \end{equation} where we understand $\boldsymbol{a}\boldsymbol{b}=\boldsymbol{a}\cdot\boldsymbol{b}$ for vectors $\boldsymbol{a}$ and $\boldsymbol{b}$. Hence, if Alice and Bob compare their results, they will find that they correlate in an oscillatory manner according to the relation above. This time-dependent correlation is merely due to entanglement, as for initial states in the product form $|\Psi_{0}\rangle=|0\rangle|1\rangle$ or $|\Psi_{0}\rangle=|1\rangle|0\rangle$ the correlation function would yield \begin{equation} \langle\hat{\boldsymbol{x}}_{A}\hat{\boldsymbol{x}}_{B}\rangle(t_{1},t_{2})=\langle0|\hat{\boldsymbol{x}}|0\rangle\langle1|\hat{\boldsymbol{x}}|1\rangle. \end{equation} Note that for the equal-time condition $t_{1}=t_{2}=t$, as well as for the periodic cases $t_{2}=t_{1}+kT$, with $k\in\{1,2,\ldots\}$, and \begin{equation} T=\frac{2\pi}{\varDelta E},\label{eq:T} \end{equation} one obtains from \eqref{eq:twotimeQM} \begin{align} \langle\hat{\boldsymbol{x}}_{A}\hat{\boldsymbol{x}}_{B}\rangle(t,t+kT) & =|\langle0|\hat{\boldsymbol{x}}|1\rangle|^{2}+\langle0|\hat{\boldsymbol{x}}|0\rangle\langle1|\hat{\boldsymbol{x}}|1\rangle\\ & =\langle\hat{\boldsymbol{x}}_{A}\hat{\boldsymbol{x}}_{B}\rangle, \end{align} where $\langle\hat{\boldsymbol{x}}_{A}\hat{\boldsymbol{x}}_{B}\rangle=\langle\Psi_{0}|(\hat{\boldsymbol{x}}\otimes\hat{\boldsymbol{x}})|\Psi_{0}\rangle$ is the single-time correlation function at initial time $t=0$.

\subsection{Analysis in Bohmian mechanics}

Let us provide an analysis of the scenario in the framework of BM and calculate the two-time correlation function of the ``true'' particle positions (which do not exist in SQM). See the Appendix for details about the notation and the foundations of BM.

The initial quantum state is given by the wavefunction \begin{equation} \Psi_{0}=\frac{1}{\sqrt{2}}\left[\phi_{0}\otimes\phi_{1}+\phi_{1}\otimes\phi_{0}\right], \end{equation} where $\phi_{n}$ are eigenstates of the local Hamiltonian $\hat{H}$, so $\hat{H}\phi_{n}=E_{n}\phi_{n}$.

The positions of the particles at time $t$ are precisely determined by the trajectory function $\xi_{t}=\int_{0}^{t}dt'(j_{t'}/\rho_{t'})$, so the time dependent density $\rho_{t}=|\Psi_{t}|^{2}$ can also be written as \begin{equation} \rho_{t}(q)=\int dq'\,\rho_{0}(q')\delta(q-\xi_{t}(q')),\label{eq:densitytraj} \end{equation} where $\rho_{0}=|\Psi_{0}|^{2}$ is the initial density at time $t=0$. For two particles $A$ and $B$, we have $q=(\boldsymbol{x},\boldsymbol{y})$, and so \eqref{eq:densitytraj} becomes \begin{equation} \rho_{t}(\boldsymbol{x},\boldsymbol{y})=\int d^{3}x'\int d^{3}y'\,\rho_{0}(\boldsymbol{x}',\boldsymbol{y}')\delta(\boldsymbol{x}-\boldsymbol{\xi}_{A,t}(\boldsymbol{x}',\boldsymbol{y}'))\delta(\boldsymbol{y}-\boldsymbol{\xi}_{B,t}(\boldsymbol{x}',\boldsymbol{y}')), \end{equation} where $\xi_{t}=(\boldsymbol{\xi}_{A,t},\boldsymbol{\xi}_{B,t})$. From the above expression it is straightforward to construct the \emph{two-time density} for the two particles, \begin{equation} \rho_{t_{1},t_{2}}(\boldsymbol{x},\boldsymbol{y})=\int d^{3}x'\int d^{3}y'\,\rho_{0}(\boldsymbol{x}',\boldsymbol{y}')\delta(\boldsymbol{x}-\boldsymbol{\xi}_{A,t_{1}}(\boldsymbol{x}',\boldsymbol{y}'))\delta(\boldsymbol{y}-\boldsymbol{\xi}_{B,t_{2}}(\boldsymbol{x}',\boldsymbol{y}')),\label{eq:twotimedensity} \end{equation} from where we can calculate the two-time correlation function of the position of the two particles as \begin{equation} \langle\boldsymbol{X}_{A}\boldsymbol{X}_{B}\rangle(t_{1},t_{2})=\int d^{3}x\int\,d^{3}y\,\boldsymbol{x}\boldsymbol{y}\,\rho_{t_{1},t_{2}}(\boldsymbol{x},\boldsymbol{y}).\label{eq:twotimerho} \end{equation} Since the initial state \eqref{eq:initialstate} is an energy eigenstate, the time evolution function becomes time-independent, $\xi_{t}(\boldsymbol{x},\boldsymbol{y})=\xi_{0}(\boldsymbol{x},\boldsymbol{y})=(\boldsymbol{x},\boldsymbol{y})$, and so the two-time density \eqref{eq:twotimedensity} becomes \begin{align} \rho_{t_{1},t_{2}}(\boldsymbol{x},\boldsymbol{y}) & =\int d^{3}x'\int d^{3}y'\,\rho_{0}(\boldsymbol{x}',\boldsymbol{y}')\delta(\boldsymbol{x}-\boldsymbol{x}')\delta(\boldsymbol{y}-\boldsymbol{y}')\\ & =\rho_{0}(\boldsymbol{x},\boldsymbol{y}). \end{align} Thus the two-time correlation function \eqref{eq:twotimerho} becomes \begin{align} \langle\boldsymbol{X}_{A}\boldsymbol{X}_{B}\rangle(t_{1},t_{2}) & =\int d^{3}x\int\,d^{3}y\,\boldsymbol{x}\boldsymbol{y}\,|\Psi_{0}(\boldsymbol{x},\boldsymbol{y})|^{2}\\ & =|\phi_{0}^{\dagger}\,\hat{\boldsymbol{x}}\,\phi_{1}|^{2}+(\phi_{0}^{\dagger}\,\hat{\boldsymbol{x}}\,\phi_{0})(\phi_{1}^{\dagger}\,\hat{\boldsymbol{x}}\,\phi_{1}),\label{eq:twotimeBMX} \end{align} where we have used that \begin{equation} \psi^{\dagger}\hat{\boldsymbol{x}}\phi=\int d^{3}x\,\boldsymbol{x}\,\psi^{*}(\boldsymbol{x})\phi(\boldsymbol{x}). \end{equation} The result coincides with the prediction \eqref{eq:twotimeQM} of standard quantum mechanics only for the equal-time condition $t_{1}=t_{2}=t$ and for periodic intervals $t_{2}=t_{1}+kT$, with $k\in\{1,2,\ldots\}$ and $T$ given by \eqref{eq:T}. For almost all other time points we have \begin{equation} \langle\boldsymbol{X}_{A}\boldsymbol{X}_{B}\rangle(t_{1},t_{2})\neq\langle\hat{\boldsymbol{x}}_{A}\hat{\boldsymbol{x}}_{B}\rangle(t_{1},t_{2}). \end{equation} So it seems that in spite of Bohm's proof of the empirical equivalence of BM and SQM, both theories do not yield the same predictions here. Furthermore, Alice and Bob can decide upon the empirical validity of either standard QM or Bohmian mechanics by repeatedly measuring at two distinct times the position of two particles initially prepared in the entangled energy eigenstate \eqref{eq:initialstate}, communicating their results and calculating the two-time correlation function.

\subsection{The hidden collapse}

The origin of the discrepancy between the predictions of SQM and of BM is that they are really not about the same quantities. The position operators $\hat{\boldsymbol{x}}_{A}$ and $\hat{\boldsymbol{x}}_{B}$ represent \emph{measured} positions, while the variables $\boldsymbol{X}_{A}$ and $\boldsymbol{X}_{B}$ represent \emph{unmeasured} positions. In the course of each measurement the state of the system is altered and hence a subsequent measurement does not necessarily yield values that coincide with those the system would possess if no preceding measurement would have been performed. Thus, comparing the expressions \eqref{eq:twotimeQM} and \eqref{eq:twotimerho} is really comparing apples with pears. In both of the above analyses carried out in the respective frameworks of SQM and BM, the measurement process has not been included as an ordinary physical process corresponding to a unitary operation on the quantum state of the system. However, on using operator-observables in the SQM analysis, we have \emph{tacitly} included the measurement process as an external intervention implying a hidden collapse of the wavefunction. We shall now analyze once more the scenario described above in the framework of SQM and reveal the hidden collapse in the calculation of the two-time correlation function \eqref{eq:twotimeAB} between arbitrary discrete observables $\hat{A}$ and $\hat{B}$. The result can then be extended, with some caution, to continuous observables, in particular to position observables.

Let us write two discrete observables $\hat{A}$ and $\hat{B}$ in their spectral decomposition, so $\hat{A}=\sum_{a}a\,\hat{\varPi}_{a}$ and $\hat{B}=\sum_{b}b\,\hat{\varPi}_{b}$, with the projections $\hat{\varPi}_{a}$ and $\hat{\varPi}_{b}$ onto the eigenspaces belonging to the eigenvalues $a$ and $b$, respectively. We then write the two-time correlation function \eqref{eq:twotimeAB} as \begin{align} \langle\hat{A}_{A}\hat{B}_{B}\rangle(t_{1},t_{2}) & =\langle\Psi|(\hat{A}(t_{1})\otimes\hat{B}(t_{2}))|\Psi\rangle\\ & =\sum_{a,b}\,ab\,\langle\Psi|\left(\hat{\varPi}_{a}(t_{1})\otimes\hat{\varPi}_{b}(t_{2})\right)|\Psi\rangle. \end{align} Using some basic operator algebra we arrive at \begin{equation} \langle\Psi|\left(\hat{\varPi}_{a}(t_{1})\otimes\hat{\varPi}_{b}(t_{2})\right)|\Psi\rangle=P_{t_{2},t_{1}}(b|a)P_{t_{1}}(a),\label{eq:expecy} \end{equation} where \begin{align} P_{t_{1}}(a) & =\langle\Psi_{t_{1}}|(\hat{\varPi}_{a}\otimes\mathbbm1)|\Psi_{t_{1}}\rangle,\\ |\Psi_{t_{1}}\rangle & =\hat{U}(t_{1})|\Psi\rangle,\\ P_{t_{2},t_{1}}(b|a) & =\langle\Psi_{t_{1},a,t_{2}}|(\mathbbm1\otimes\hat{\varPi}_{b})|\Psi_{t_{1},a,t_{2}}\rangle,\\ |\Psi_{t_{1},a,t_{2}}\rangle & =\hat{U}(t_{2}-t_{1})\frac{1}{\sqrt{P_{t_{1}}(a)}}(\hat{\varPi}_{a}\otimes\mathbbm1)\hat{U}(t_{1})|\Psi\rangle. \end{align} The expression $P_{t_{1}}(a)$ represents the probability that Alice finds her particle at time $t_{1}$ having the property $a$. For $t_{1}<t_{2}$ the vector $|\Psi_{t_{1},a,t_{2}}\rangle$ can be interpreted as the state resulting from the following sequence of operations: The initial state $|\Psi\rangle$ freely evolves up to time $t_{1}$, is then projected onto the eigenspace of $a$ and subsequently renormalized, and then again freely evolves up to time $t_{2}$. These operations correspond to Alice detecting at time $t_{1}$ her particle with the property $a$. Hence, the function $P_{t_{2},t_{1}}(b|a)$ represents the conditional probability that Bob finds his particle at time $t_{2}$ having property $b$ given that Alice previously had found her particle at time $t_{1}$ having property $a$. Consequently, the righthand side of \eqref{eq:expecy} can be understood as the joint probability $P_{t_{1},t_{2}}(a,b)$ that Alice finds her particle at earlier time $t_{1}$ having property $a$, and Bob finds his particle at time $t_{2}$ having property $b$, so that \begin{equation} P_{t_{2},t_{1}}(b|a)P_{t_{1}}(a)=P_{t_{1},t_{2}}(a,b). \end{equation} Now, since the measurements are performed on different subsystems, they commute, and thus the temporal ordering of the measurements actually does not matter. For temporally reversed measurements $t_{2}<t_{1}$, calculations analog to those above reveal that \begin{equation} \langle\Psi|\left(\hat{\varPi}_{a}(t_{1})\otimes\hat{\varPi}_{b}(t_{2})\right)|\Psi\rangle=P_{t_{2}}(b)P_{t_{1},t_{2}}(a|b),\label{eq:expecx} \end{equation} with \begin{align} P_{t_{2}}(b) & =\langle\Psi_{t_{2}}|(\mathbbm1\otimes\hat{\varPi}_{b})|\Psi_{t_{2}}\rangle,\\ |\Psi_{t_{2}}\rangle & =\hat{U}(t_{2})|\Psi\rangle,\\ P_{t_{1},t_{2}}(b|a) & =\langle\Psi_{t_{2},b,t_{1}}|(\hat{\varPi}_{a}\otimes\mathbbm1)|\Psi_{t_{2},b,t_{1}}\rangle,\\ |\Psi_{t_{2},b,t_{1}}\rangle & =\hat{U}(t_{1}-t_{2})\frac{1}{\sqrt{P_{t_{2}}(b)}}(\mathbbm1\otimes\hat{\varPi}_{b})\hat{U}(t_{2})|\Psi\rangle. \end{align} The expression $P_{t_{2}}(b)$ represents the probability that Bob finds his particle at time $t_{2}$ having property $b$. The vector $|\Psi_{t_{2},b,t_{1}}\rangle$ can be interpreted as the state resulting from a sequence of operations reverse to those given further above: Bob first performs a measurement on his particle at time $t_{2}$ and finds it having property $b$, then the state freely evolves up to time $t_{1}$. Hence, the function $P_{t_{1},t_{2}}(a|b)$ represents the conditional probability that Alice finds her particle at time $t_{1}$ having property $a$ under the condition that Bob previously had found his particle at time $t_{2}$ having property $b$. Consequently, also the righthand side of \eqref{eq:expecx} can be understood as the joint probability $P_{t_{1},t_{2}}(a,b)$ that Bob finds his particle at time $t_{2}$ having property $a$ and Alice finds her particle at earlier time $t_{1}$ having property $b$, so that \begin{equation} P_{t_{1},t_{2}}(a|b)P_{t_{2}}(b)=P_{t_{1},t_{2}}(a,b). \end{equation} Lastly, on the equal-time condition $t_{1}=t_{2}=t$ we have \begin{equation} \hat{\varPi}_{a}(t)\otimes\hat{\varPi}_{b}(t)=\hat{U}^{\dagger}(t)(\hat{\varPi}_{a}\otimes\hat{\varPi}_{b})\hat{U}(t), \end{equation} and thus we directly obtain the joint probability that Alice and Bob find their particles at time $t$ having properties $a$ and $b$, respectively, \begin{equation} \langle\Psi|\left(\hat{\varPi}_{a}(t)\otimes\hat{\varPi}_{b}(t)\right)|\Psi\rangle=P_{t}(a,b).\label{eq:expecxt} \end{equation} The above calculations have been carried out using operators with a discrete spectrum, so what about continuous observables like position? The spectral decomposition of the position operator, \begin{equation} \hat{\boldsymbol{x}}=\int d^{3}x\,\boldsymbol{x}\,\hat{\varPi}_{x}, \end{equation} involves the improper projections $\hat{\varPi}_{x}=|\boldsymbol{x}\rangle\langle\boldsymbol{x}|$, which do not fulfill the crucial property of idempotency, $\hat{\varPi}_{x}^{2}=\hat{\varPi}_{x}$, but rather a generalized version of it, $\hat{\varPi}_{x}\hat{\varPi}_{x'}=\hat{\varPi}_{x}\delta(\boldsymbol{x}-\boldsymbol{x}')$. This reflects the fact that continuous observables are really idealizations. In practice it is not possible to perform an exact measurement of a continuous observable like position. Instead, position would be measured by, say, an array of detectors, each one indicating whether or not the particle is located within a small but finite spatial region $X_{i}\subset\mathbbm R^{3}$, so that the corresponding operator would have a discrete spectrum and can be decomposed as \begin{equation} \hat{\boldsymbol{x}}_{\varDelta}=\sum_{i}\boldsymbol{x}_{i}\,\hat{\varPi}_{i}, \end{equation} where the $\boldsymbol{x}_{i}$ are the centers of the regions $X_{i}$, where $\varDelta$ corresponds to the spatial separation of the centers, and where \begin{equation} \hat{\varPi}_{i}=\int_{X_{i}}d^{3}x\,\hat{\varPi}_{x} \end{equation} are projections onto the respective regions $X_{i}$. In the theoretical limit $\varDelta\rightarrow0$ one would obtain the exact position operator $\hat{\boldsymbol{x}}_{\varDelta}\rightarrow\hat{\boldsymbol{x}}$, but in practice $\varDelta$ is bounded from below due to technological limitations. With these precautions in mind we can extend the results of the previous section, which were obtained for discrete observables $\hat{A}$ and $\hat{B}$, to the idealized case of continuous observables like position.

To summarize, for distinct times $t_{1}\neq t_{2}$, irrespective of the temporal ordering, we have seen that there is a hidden collapse built into the two-time correlation function obtained in SQM, and this collapse generates a correlation depending on the time interval between the two measurements. Since the analysis carried out in the framework of BM using the position variables $\boldsymbol{X}_{A}$ and $\boldsymbol{X}_{B}$ did not include any measurement, and therefore no alteration of the system state induced by measurement, the obtained two-time correlation function cannot be expected to coincide with the two-time correlation function obtained in SQM using operators $\hat{\boldsymbol{x}}_{A}$ and $\hat{\boldsymbol{x}}_{B}$. This is not just a mathematical subtlety. The two-time correlation function obtained in SQM involves \emph{measured} values, while the two-time correlation function obtained in BM involves \emph{unmeasured} values, that is, values for the position of the particles \emph{without} the disturbance of a measurement. On the equal-time condition the two-time correlation functions obtained in SQM and BM coincide. The reason for this coincidence is that when Alice and Bob measure their particles at the same time, the collapses introduced by their measurements fall together to one single collapse. Hence, there can be no influence of the collapse introduced by the measurement of one party on the measurement result of the other party.

\subsection{Re-analysis in Bohmian mechanics including measurement}

According to the considerations of the previous section, BM should yield the same predictions as SQM when position measurements by Alice and Bob are included in the analysis. Again, let us perform the re-analysis at first with discrete observables. The results may then be extended, with the already mentioned precautions, to the case of continuous observables like position. See the Appendix for a general treatment of measurement in BM.

Alice and Bob measure discrete observables $\hat{A}$ and $\hat{B}$ at times $t_{1}$ and $t_{2}$, respectively, by using measurement devices $M_{A}$ and $M_{B}$ having pointer states $\eta$ and $\mu$, respectively. Hence, the total system is initially in the state \begin{equation} \Psi_{0}=\psi_{0}\otimes\eta_{0}\otimes\mu_{0}, \end{equation} where \begin{equation} \psi_{0}=\frac{1}{\sqrt{2}}\left[\phi_{0}\otimes\phi_{1}+\phi_{1}\otimes\phi_{0}\right]\label{eq:initialBohm} \end{equation} is the initial state of the system of interest already introduced in \eqref{eq:initialstate}, and $\eta_{0},\mu_{0}$ are initial pointer states. Alice and Bob want to measure the observables $\hat{A}$ and $\hat{B}$ at distinct times $t_{1}$ and $t_{2}$, respectively, where $t_{1}<t_{2}$.

The system freely evolves from $t=0$ up to the time $t_{1}-T_{M}$ when Alice's measurement begins, with $T_{M}$ being a small but finite time interval representing the duration of the measurement, so that the free evolution of the system during a period of length $T_{M}$ can be neglected. The state immediately before Alice's measurement then reads \begin{equation} \Psi_{t_{1}-T_{M}}=\psi_{t_{1}-T_{M}}\otimes\eta_{t_{1}-T_{M}}\otimes\mu_{t_{1}-T_{M}}, \end{equation} with \begin{align} \psi_{t_{1}-T_{M}} & =e^{-i\hat{H}_{AB}(t_{1}-T_{M})}\psi_{0}\\ & =e^{-i(E_{0}+E_{1})(t_{1}-T_{M})}\psi_{0}, \end{align} As for the pointer state $\eta$, it must have been prepared in such a way that when the measurement begins it is in the ``ready'' state $\eta_{R}$, so that $\eta_{t_{1}-T_{M}}=\eta_{R}$, and therefore \begin{equation} \Psi_{t_{1}-T_{M}}=\psi_{0}\otimes\eta_{R}\otimes\mu_{t_{1}-T_{M}}, \end{equation} where we have removed the irrelevant global phase factor $e^{-i(E_{0}+E_{1})(t_{1}-T_{M})}$.

Now Alice performs her measurement of the observable $\hat{A}$, resulting in the post-measurement state \begin{equation} \Psi_{t_{1}}=\sum_{a}\psi_{a}\otimes\eta_{a}\otimes\mu_{t_{1}-T_{M}}, \end{equation} where $\psi_{a}=(\hat{\varPi}_{a}\otimes\mathbbm1)\psi_{0}$ are unnormalized eigenvectors of $\hat{A}$ with corresponding eigenvalues $a$, and with $\sum_{a}\|\psi_{a}\|^{2}=1$. The pointer states $\eta_{a}$ corresponding to different outcomes ``$a$'' have by construction (see Appendix) approximately zero overlap, so that for $a\neq a'$ \begin{align} \eta_{a}(z_{A})\eta_{a'}(z_{A}) & \approx0, \end{align} for all configurations $z_{A}$ of Alice's measurement device. Furthermore, the pointer states $\eta_{a}$ have by construction (see Appendix) almost all of their support within the respective regions $Z_{a}$ corresponding to the outcomes ``$a$'' , so \begin{align} \int_{Z_{a}}dz_{a}|\eta_{a}(z_{A})|^{2} & \approx1. \end{align} These relations have to be kept in mind for later use. Now, the system freely evolves up to a later time $t_{2}-T_{M}$ when Bob's measurement begins, reaching the state \begin{equation} \Psi_{t_{2}-T_{M}}=\sum_{a}\psi_{t_{1},a,t_{2}-T_{M}}\otimes\eta_{t_{1},a,t_{2}-T_{M}}\otimes\mu_{R}, \end{equation} with \begin{align} \psi_{t_{1},a,t_{2}-T_{M}} & =e^{-i\hat{H}_{AB}(t_{2}-T_{M}-t_{1})}\psi_{a}\\ & \approx e^{-i(\hat{H}_{A}+\hat{H}_{B})(t_{2}-t_{1})}(\hat{\varPi}_{a}\otimes\mathbbm1)\psi_{0}\\ & =\left(e^{-i\hat{H}_{A}(t_{2}-t_{1})}\hat{\varPi}_{a}\otimes\mathbbm1\right)\frac{1}{\sqrt{2}}\left(e^{-iE_{1}(t_{2}-t_{1})}\phi_{0}\phi_{1}+e^{-iE_{0}(t_{2}-t_{1})}\phi_{1}\phi_{0}\right), \end{align} where we have exploited the shortness of the measurement duration $T_{M}$. As for the pointer states we have \begin{align} \eta_{t_{1},a,t_{2}-T_{M}} & =e^{-i\hat{H}_{M_{A}}(t_{2}-T_{M}-t_{1})}\eta_{R},\\ \mu_{R} & =e^{-i\hat{H}_{M_{B}}(t_{2}-T_{M})}\mu_{0}, \end{align} where we have used that Bob's pointer state must have been prepared in such a way that before his measurement begins, the pointer is in the ``ready'' state $\mu_{R}$. Now Bob performs his measurement of the observable $\hat{B}$ resulting in the post-measurement state \begin{equation} \Psi_{t_{2}}=\sum_{a,b}\psi_{t_{1},a,t_{2},b}\otimes\eta_{t_{2},a}\otimes\mu_{b},\label{eq:psiuncollapsed} \end{equation} where \begin{align} \psi_{t_{1},a,t_{2},b} & =(\mathbbm1\otimes\hat{\varPi}_{b})\psi_{t_{1},a,t_{2}-T_{M}}\\ & =\left(e^{-i\hat{H}_{A}(t_{2}-t_{1})}\hat{\varPi}_{a}\otimes\hat{\varPi}_{b}\right)\frac{1}{\sqrt{2}}\left(e^{-iE_{1}(t_{2}-t_{1})}\phi_{0}\phi_{1}+e^{-iE_{0}(t_{2}-t_{1})}\phi_{1}\phi_{0}\right)\label{eq:psitfinal} \end{align} are unnormalized eigenvectors of both $\hat{A}$ and $\hat{B}$ with corresponding eigenvalues $a$ and $b$, respectively. The state \eqref{eq:psiuncollapsed} is an uncollapsed state resulting from a continuous unitary evolution describing the measurements of both Alice and Bob, as well as the free evolution between the measurements. The resulting state is a sum of branches, \begin{align} \Psi_{t_{2}} & =\sum_{a,b}\Psi_{t_{1},a,t_{2},b}, \end{align} where each branch corresponds to a different combination of outcomes of Alice's and Bob's measurements, \begin{equation} \Psi_{t_{1},a,t_{2},b}=\psi_{t_{1},a,t_{2},b}\otimes\eta_{a,t_{2}}\otimes\mu_{b}. \end{equation} The pointer states $\mu_{b}$ corresponding to different outcomes ``$b$'' have by construction approximately zero overlap, and they have almost all of their support within the respective regions $Z_{b}$ corresponding to the respective outcomes ``$b$''. Now, as each macroscopic measurement device involves a large number of particles, the huge number of internal degrees of freedom induces a further rapid decrease of the overlap between the time-evolved pointer states $\eta_{a,t_{2}}=\hat{U}_{M_{A}}(t_{2}-t_{1})\eta_{a}$. The same argument is used in\emph{ }decoherence theory to explain why branches of the wavefunction that are coupled to an external macroscopic reservoir decohere. As a consequence of such decoherence, the branches stay approximately orthogonal to each other for all temporal distances $t_{2}-t_{1}>0$, so for $a\neq a'$ and $b\neq b'$ we have \begin{align} \Psi_{t_{1},a,t_{2},b}^{*}(q)\Psi_{t_{1},a',t_{2},b'}(q) & \approx0\label{eq:orthog} \end{align} for all configurations $q$, and therefore \begin{equation} |\Psi_{t_{2}}(q)|^{2}\approx\sum_{a,b}|\Psi_{t_{1},a,t_{2},b}(q)|^{2}.\label{eq:Psit2} \end{equation} Now consider the following history of events. Say, at time $t_{1}$ Alice obtains the result ``$a$''. Then the trajectory of the system crosses at time $t_{1}$ the region $Q_{a}=\mathcal{Q}_{S}\times Z_{a}\times\mathcal{Q}_{M_{B}}$, where $\mathcal{Q}_{S}$ is the configuration space of the system of interest, $Z_{a}$ is the region in the configuration space $\mathcal{Q}_{M_{A}}$ of Alice's measurement device corresponding to the result ``$a$'', and $\mathcal{Q}_{M_{B}}$ is the configuration space of Bob's measurement device. Thus, at the later time $t_{2}$ the trajectory crosses the region $Q_{a}=\mathcal{Q}_{S}\times Z_{a,t_{2}}\times\mathcal{Q}_{M_{B}}$, where $Z_{a,t_{2}}$ is the region obtained by propagating every point in $Z_{a}$ from time $t_{1}$ to time $t_{2}$ along its unique trajectory. We therefore have \begin{equation} \int_{Z_{a,t_{2}}}dz\,|\eta_{a,t_{2}}(z)|^{2}=\int_{Z_{a}}dz\,|\eta_{a}(z)|^{2}.\label{eq:etaint} \end{equation} Now say that Bob obtains at time $t_{2}$ the result ``$b$'', then the system trajectory crosses at time $t_{2}$ the region \begin{equation} Q_{a,b}=\mathcal{Q}_{S}\times Z_{a,t_{2}}\times Z_{b}, \end{equation} where $Z_{b}$ is the region in the configuration space of Bob's measurement device corresponding to the result ``$b$''. The joint probability for the occurrence of the outcomes $a$ and $b$ at the times $t_{1}$ and $t_{2}$, respectively, thus reads \begin{align} P_{t_{1},t_{2}}(a,b) & =\int_{Q_{a,b}}dq\,|\Psi_{t_{2}}(q)|^{2}\\ & \approx\int_{Q_{a,b}}dq\sum_{a',b'}|\Psi_{t_{1},a',t_{2},b'}(q)|^{2}\\ & \approx\int_{Q_{a,b}}dq\,|\Psi_{t_{1},a,t_{2},b}(q)|^{2}\\ & =\int d^{3}x\int d^{3}y\,|\psi_{t_{1},a,t_{2},b}(\boldsymbol{x},\boldsymbol{y})|^{2}\int_{Z_{a,t_{2}}}dz_{A}|\eta_{a,t_{2}}(z_{A})|^{2}\int_{Z_{b}}dz_{B}|\mu_{b}(z_{B})|^{2}\\ & =\int d^{3}x\int d^{3}y\,|\psi_{t_{1},a,t_{2},b}(\boldsymbol{x},\boldsymbol{y})|^{2}\int_{Z_{a}}dz_{A}|\eta_{a}(z_{A})|^{2}\int_{Z_{b}}dz_{B}|\mu_{b}(z_{B})|^{2}\\ & \approx\int d^{3}x\int d^{3}y\,|\psi_{t_{1},a,t_{2},b}(\boldsymbol{x},\boldsymbol{y})|^{2}, \end{align} where we have used \eqref{eq:Psit2} and \eqref{eq:etaint}, as well as the fact that the pointer states $\eta_{a}$ and $\mu_{b}$ have almost all of their support in $Z_{a}$ and $Z_{b}$, respectively. Using \eqref{eq:initialBohm} and \eqref{eq:psitfinal} we obtain \begin{equation} \begin{split}P_{t_{1},t_{2}}(a,b) & \approx\Re\left\{ (\phi_{0}^{\dagger}\hat{\varPi}_{b}\phi_{1})(\phi_{1}^{\dagger}\hat{\varPi}_{b}\phi_{0})e^{i(E_{1}-E_{0})(t_{2}-t_{1})}\right\} \\ & \phantom{=}+\frac{1}{2}\left[(\phi_{0}^{\dagger}\hat{\varPi}_{a}\phi_{0})(\phi_{1}^{\dagger}\hat{\varPi}_{b}\phi_{1})+(\phi_{1}^{\dagger}\hat{\varPi}_{a}\phi_{1})(\phi_{0}^{\dagger}\hat{\varPi}_{b}\phi_{0})\right]. \end{split} \label{eq:twotimeP} \end{equation} Thus, the two-time correlation function of the two operators $\hat{A}$ and \textbf{$\hat{B}$} obtained by Bohmian mechanics reads \begin{align} \langle\hat{A}\hat{B}\rangle(t_{1},t_{2}) & =\sum_{a,b}ab\,P_{t_{1},t_{2}}(a,b)\\ & \begin{aligned}\approx\, & \Re\left\{ (\phi_{0}^{\dagger}\hat{A}\phi_{1})(\phi_{1}^{\dagger}\hat{B}\phi_{0})e^{i(E_{1}-E_{0})(t_{2}-t_{1})}\right\} \\ & +\frac{1}{2}\left[(\phi_{0}^{\dagger}\hat{A}\phi_{0})(\phi_{1}^{\dagger}\hat{B}\phi_{1})+(\phi_{1}^{\dagger}\hat{A}\phi_{1})(\phi_{0}^{\dagger}\hat{B}\phi_{0})\right], \end{aligned} \end{align} which approximately coincides with the prediction \eqref{eq:twotimeAB} of standard quantum mechanics. We may then extend this result, with the usual precautions mentioned further above, to continuous observables, so that for $\hat{A}=\hat{B}=\hat{\boldsymbol{x}}$ we obtain \begin{align} & \begin{aligned}\langle\hat{\boldsymbol{x}}_{A}\hat{\boldsymbol{x}}_{B}\rangle(t_{1},t_{2})\approx\, & |\phi_{0}^{\dagger}\hat{\boldsymbol{x}}\phi_{1}|^{2}\cos\left(\varDelta E(t_{2}-t_{1})\right)\\ & +(\phi_{0}^{\dagger}\hat{\boldsymbol{x}}\phi_{0})(\phi_{1}^{\dagger}\hat{\boldsymbol{x}}\phi_{1}), \end{aligned} \label{eq:twotimeBM} \end{align} which approximately coincides with the prediction \eqref{eq:twotimeQM} of SQM.

Formally, the final result \eqref{eq:twotimeBM} can be evaluated for any combination of values for $t_{1}$ and $t_{2}$. However, the derivation above requires the temporal ordering $t_{1}<t_{2}$. As can easily be verified, an analog calculation carried out with reversed temporal ordering $t_{2}<t_{1}$ leads to the same end result \eqref{eq:twotimeBM}. Finally, let us give a derivation on the equal-time condition $t_{1}=t_{2}=t$. Alice and Bob simultaneously measure their observables at time $t$, so the quantum state of the system immediately before their measurement reads \begin{equation} \Psi_{t-T_{M}}=\psi_{0}\otimes\eta_{R}\otimes\mu_{R}. \end{equation} Immediately after their measurement, the state becomes \begin{equation} \Psi_{t}=\sum_{a,b}\psi_{a,b}\otimes\eta_{a}\otimes\mu_{b}, \end{equation} where \begin{equation} \psi_{a,b}=(\hat{\varPi}_{a}\otimes\hat{\varPi}_{b})\psi_{0}.\label{eq:psitfinal2} \end{equation} Hence, the resulting state is a sum of branches, \begin{align} \Psi_{t} & =\sum_{a,b}\Psi_{a,b}, \end{align} where each branch corresponds to a different combination of outcomes of Alice's and Bob's measurements, \begin{equation} \Psi_{a,b}=\psi_{a,b}\otimes\eta_{a}\otimes\mu_{b}. \end{equation} The pointer states $\eta_{a}$ and $\mu_{b}$ have each by construction approximately zero overlap for different $a$ and $b$, respectively, so the branches are approximately orthogonal to each other, thus for $a\neq a'$ and $b\neq b'$ we have \begin{align} \Psi_{a,b}^{*}(q)\Psi_{a',b'}(q) & \approx0,\label{eq:orthog-1} \end{align} and therefore \begin{equation} |\Psi_{t}(q)|^{2}\approx\sum_{a,b}|\Psi_{a,b}(q)|^{2}.\label{eq:Psit2-1} \end{equation} Say at time $t$ Alice obtains the result ``$a$'' and Bob obtains the result ``$b$''. Then the trajectory of the system crosses at time $t$ the region $Q_{a,b}=\mathcal{Q}_{S}\times Z_{a}\times Z_{b}$. Calculations analog to those carried out further above yields the probability for the joint occurrence of the measurement results ``$a$'' and ``$b$'' as \begin{equation} P_{t}(a,b)\approx\int d^{3}x\int d^{3}y\,|\psi_{a,b}(\boldsymbol{x},\boldsymbol{y})|^{2}. \end{equation} For the initial state \eqref{eq:initialBohm} and using \eqref{eq:psitfinal2} we obtain \begin{equation} \begin{split}P_{t}(a,b) & \approx\Re\left\{ (\phi_{0}^{\dagger}\hat{\varPi}_{a}\phi_{1})(\phi_{1}^{\dagger}\hat{\varPi}_{b}\phi_{0})\right\} \\ & \phantom{=}+\frac{1}{2}\left[(\phi_{0}^{\dagger}\hat{\varPi}_{a}\phi_{0})(\phi_{1}^{\dagger}\hat{\varPi}_{b}\phi_{1})+(\phi_{1}^{\dagger}\hat{\varPi}_{a}\phi_{1})(\phi_{0}^{\dagger}\hat{\varPi}_{b}\phi_{0})\right]. \end{split} \label{eq:twotimeBM-1} \end{equation} Thus, the two-time correlation function of the two operators $\hat{A}$ and \textbf{$\hat{B}$} obtained by Bohmian mechanics on the equal-time condition $t_{1}=t_{2}=t$ reads \begin{align} \langle\hat{A}\hat{B}\rangle(t,t) & =\sum_{a,b}ab\,P_{t}(a,b)\\ & \begin{aligned}\approx\, & \Re\left\{ (\phi_{0}^{\dagger}\hat{A}\phi_{1})(\phi_{1}^{\dagger}\hat{B}\phi_{0})\right\} \\ & +\frac{1}{2}\left[(\phi_{0}^{\dagger}\hat{A}\phi_{0})(\phi_{1}^{\dagger}\hat{B}\phi_{1})+(\phi_{1}^{\dagger}\hat{A}\phi_{1})(\phi_{0}^{\dagger}\hat{B}\phi_{0})\right], \end{aligned} \end{align} which approximately coincides with the prediction \eqref{eq:twotimeAB} of standard quantum mechanics for equal times $t_{1}=t_{2}=t$. Again replacing the discrete observables $\hat{A}$ and $\hat{B}$ with the continuous position operator $\hat{\boldsymbol{x}}$, we obtain \begin{equation} \begin{split}\langle\hat{\boldsymbol{x}}_{A}\hat{\boldsymbol{x}}_{B}\rangle(t,t) & \approx\,|\phi_{0}^{\dagger}\hat{\boldsymbol{x}}\phi_{1}|^{2}+(\phi_{0}^{\dagger}\hat{\boldsymbol{x}}\phi_{0})(\phi_{1}^{\dagger}\hat{\boldsymbol{x}}\phi_{1}),\end{split} \label{eq:equaltimeBM} \end{equation} which approximately coincides with the prediction \eqref{eq:twotimeQM} for equal times.

Concluding, for all temporal distances and orderings between the measurements of Alice and Bob we obtain approximately the same predictions for the two-time correlation function in BM that has been obtained in SQM using operator algebra. The quality of the approximation depends on the spatial separation between the wave packets of the pointer states corresponding to the different measurement outcomes. In the idealized case of perfect separation the predictions of SQM and BM fully coincide. The measurement process takes a small but finite amount of time, so there is no abrupt alteration of the quantum state, and since the particles are guided by the wavefunction there is neither an abrupt alteration of the particle trajectories. Note that in the analysis we did not make use of the \emph{effective collapse}, which is a mathematically convenient, but not essential, part of the Bohmian methodology. Instead, we used the uncollapsed wavefunction and evaluated the final expression for the joint probability of two-time measurement outcomes with the associated history of events.

\section{Discussion}

I have shown that if the measurement process is adequately accounted for, Bohmian mechanics is able to approximately reproduce the predictions of standard quantum mechanics also for the case of two-time correlation functions involving entangled states. The fact that these are only approximations does not have to be regarded as a flaw of Bohmian mechanics but can also be viewed as a gain in realism. The quality of the approximation depends on the spatial separation between the wave packets of the pointer states corresponding to the different measurement outcomes. On the other hand, the operator algebra used in SQM tacitly assumes a perfect ability to distinguish between the eigenvalues of the measured operator. If the separation of the pointer wave packets could be made perfect then the predictions of BM would fully coincide with those of SQM. However, perfectly separated wave packets would require the pointer wavefunctions to have only finite support on configuration space, and such wavefunctions typically have infinite average kinetic energy due to discontinuities of the wavefunctions and its derivative at the boundary, which is clearly an unrealistic scenario.

Let us explicitly go through some critical statements in the challenging paper by Kiukas and Werner and see whether the issues can be resolved. The paper starts with the statement 
\begin{quote} 
It is well-known that the position operators of a particle at different times do not in general commute. This is the reason why the notion of trajectories cannot be applied to quantum particles. 
\end{quote} 
This statement is idiosyncratic as it shows the fundamental disagreement between the adherents and opponents of BM. At the core of this fundamental disagreement lies a different stance regarding what a physical theory has to provide. For some adherents of SQM an operator-observable is simply the mathematical representation of a physical quantity of the system. Therefore, if an operator-observable evolves in time in such a way that it does not even commute with itself at a later time, then there can be no representation of the objective history of that quantity. For the Bohmian, as well as for some proponents of SQM such as Bohr and Heisenberg, an operator-observable just represents an \emph{experimental procedure} to unveil the value of a physical quantity. If this representation does not commute with itself at different times, then this only means that it is impossible to \emph{measure} the quantity from outside the system without affecting its evolution. Indeed, when the position of a particle is measured then due to Heisenberg uncertainty the momentum of the particle is altered in an uncontrollable and potentially drastic fashion, which consequently also alters the future position of the particle. A strong adherent of SQM might take a radically positivist stance by saying that there simply \emph{is} nothing beyond what can be measured. So, when the position of a particle cannot be measured without affecting the trajectory in an uncontrollable way, then there is no such thing as a trajectory. A weaker form of positivism would be to acknowledge that there \emph{might be something} beyond measurement, but to hold that it is \emph{irrelevant} for a physical theory. A defender of BM would consider both of these positivist stances unsatisfactory. To him, a physical theory must provide a complete picture of nature \emph{as it is}, and not \emph{as it appears} to some observer with a measurement device. These different views are plainly incompatible. At least, however, one may acknowledge that it is \emph{per se} neither inconsistent nor necessarily empirically false to consider a particle trajectory beyond measurement as physically real.

The authors further write: 
\begin{quote} 
In {[}Bohmian mechanics and Nelson's theory{]} the positions at all times have a joint distribution, and therefore cannot violate a Bell inequality. Hence their predictions must be in conflict with quantum mechanics and, most likely, with experiment. 
\end{quote} 
As I have tried to substantiate, the mentioned predictions are only \emph{apparently} in conflict with each other. The reason being that the predictions of SQM involve operator-observables representing \emph{measured} positions of the particles, while the seemingly analog calculation in BM involves classical variables representing the \emph{unmeasured} positions. Therefore, the two predictions are not about the same things and cannot be played off against each other. Whether or not it is reasonable or physically sound to consider such thing as an ``unmeasured position'' is not at stake here. Moreover, I have shown that the predictions of BM approximately agree with those of SQM once the measurement process has suitably been taken into account. The fact that the agreement is only approximative does not suffice to claim a conflict between the theories, as the difference between the predictions can be made arbitrarily small in theory, and in practice it is only limited by the technological advances.

Later on, Kiukas and Werner pick up, and attempt to defeat, a crucial argument of the defenders of BM: 
\begin{quote} 
The simplest position is to include the collapse of the wave function into the theory {[}citations{]}. Then the first measurement instantaneously collapses the wave function. So if agreement with quantum mechanics is to be kept, the probability distribution changes suddenly. There is no way to fit this with continuous trajectories: When the guiding field collapses, the particles must jump. While the glaring non-locality of this process may be seen as just another instance of implicate order, it introduces an element of unexplained randomness, and demotes the Bohm equation (or Nelson’s Fokker-Planck equation) from its role as the fundamental dynamical equation for position. 
\end{quote} 
Since BM is a collapse-free theory, it might be surprising to hear that the defenders of the theory involve, of all things, the \emph{collapse} to explain away conflicting predictions. In their response to the challenging article, D\"urr et al. write 
\begin{quote} 
It is easy to see that $\psi$ governs the evolution of the actual configuration $X$ of the subsystem in the usual Bohmian way, and that it collapses upon measurement of the subsystem according to the usual quantum mechanical rules, with probabilities given by the Born rule. 
\end{quote} 
Here, the authors refer to the so-called \emph{conditional wavefunction}, which is given by 
\begin{equation}
	\psi_t(x)\sim\Psi_t(x,Y_t), 
\end{equation} 
where $Y_t$ is the actual configuration of the measurement device at time $t$. The cited argument makes an appeal to the ``collapse'' at a crucial point. This collapse, however, is not the collapse as it is usually understood in the context of SQM. It is not the "Proze{\ss} 1" introduced by von Neumann in his monumental textbook  \cite{Neumann1932}: a discontinuous projection and renormalization of the wavefunction. The collapse mentioned above by defenders of BM is no sudden, indeterministic ``jump'' of neither the wavefunction nor the particles. The measurement process takes a small but finite amount of time, during which the wavefunction of the total system continuously and deterministically changes, and so do the particle positions as they are guided by the wavefunction. Discontinuity and randomness is nowhere at play in the measurement process. The probabilistic element is introduced not at the time of measurement but long before due to insufficient knowledge about the initial conditions of the entire experiment. The result of the continuous measurement process is a wavefunction with one particular branch, which is the only branch occupied by particles, which is associated with the conditional wavefunction, and which behaves just \emph{as if} it were the result of an ordinary projection of the pre-measurement wavefunction, hence the result of a ``Proze{\ss} 1''. The absolute square of this branch just equals the probability value provided by Born's rule. So, in the interpretation of Bohmian mechanics, the probability that the particles in fact \emph{do} occupy that branch, and hence the probability of actually obtaining the particular measurement result associated with that branch, is just the one predicted by SQM.

So, what is the ontological status of the conditional wavefunction? Being a factorial part of one branch of a really existing wavefunction (in the interpretation of BM), it really exists. However, for the same reason, it is not a fundamental but rather a \emph{derived} entity; in the same manner that, for example, the equator, being part of the earth, is a really existing but derived entity. There is no actual reason to consider the conditional wavefunction ``more real'' (whatever that means) than any of the other branches of the wavefunction. It is, however, more \emph{physically relevant}, as it is associated with the only branch of the wavefunction that governs the future behavior of the particles from the moment that the measurement is finished. So, it is reasonable, although not \emph{necessary}, to neglect the other branches of the wavefunction and perform all post-measurement calculations using only the conditional wavefunction. Put differently: it \emph{appears} to the observer as if all other branches have vanished and only the branch  picked out by the conditional wavefunction is left over. Since that branch coincides with a projection of the pre-measurement wavefunction on just the subspace associated with the measurement result actually obtained, we therefore face an \emph{explanation} of the seemingly discontinuous, random ``collapse'' of the wavefunction induced by a measurement process. In the framework of BM, the collapse of the wavefunction, albeit just an ``effective'' collapse involving only the physically relevant part of the wavefunction, is a theorem, not a postulate. To repeat, it is not the entire wavefunction that collapses randomly and discontinuously, but it is just the \emph{physically relevant part} of the wavefunction that collapses smoothly and deterministically. Randomness enters the description only \emph{subjectively}, due to our insufficient knowledge about what part of the wavefunction \emph{really is} is the physically relevant one.

In citing a defense strategy of the Bohmians, Kiukas and Werner try to turn the argument against the arguers: \begin{quote} So {[}Bohmians argue that{]} the two-time correlations computed from the 2-particle ensemble of trajectories are never observed anyhow, and hence pose no threat to the theory. The downside of this argument is that it also applies to single time measurements, i.e., the agreement between Bohm-Nelson configurational probabilities and quantum ones is equally irrelevant. The naive version of Bohmian theory holds “position” to be special, even “real”, while all other measurement outcomes can only be described indirectly by including the measurement devices. Saving the Nelson-Bohm theory’s failure regarding two-time two-particle correlations by going contextual also for position just means that the particle positions are declared unobservable according to the theory itself, hence truly hidden. \end{quote} The predictions in the framework of BM involving the actual positions $\boldsymbol{X}_{A}$ and $\boldsymbol{X}_{B}$ without the measurement procedures \eqref{eq:twotimeBMX}, as well as the predictions involving the measurement procedures \eqref{eq:equaltimeBM} both coincide with the predictions performed in the framework of SQM involving the position operators $\hat{\boldsymbol{x}}_{A}$ and $\hat{\boldsymbol{x}}_{B}$ \eqref{eq:twotimeQM} on the equal-time condition $t_{1}=t_{2}=t$. Thus, when Alice and Bob simultaneously measure, they find their particles at the positions where BM says they are ``truly'' located. So I find no basis for the claim that equal-time measurements pose a problem to BM and that the true particle positions are always unobservable. It is only for distinct-time measurements that the secondly measured position does not coincide with its unmeasured counterpart, and this is not surprising because the first measurement on one particle disturbs the quantum state of the total system in a nonlocal way so as to affect also the course of the trajectory of the other particle. BM is a nonlocal hidden-variables theory, and a local measurement on one part of the system may have immediate consequences on the particle trajectories in a remote part of the system. One might find this kind of nonlocal realism disturbing or even unacceptable, but if BM were a \emph{local} hidden-variables theory than it would immediately fall prey to all sorts of CHSH inequalities, and would therefore easily be shown to be empirically false. BM might be weird in its insistence on position as a ``special'' or ``real'' quantity, but as far as the here presented analysis shows, it cannot be accused of inconsistency or of empirical inadequacy. Bohmian particle positions \emph{can} be observed, though each observation has consequences on the outcome of future measurements, even for particles that are very far away.

\section{Appendix}

\subsection{Notation}

We use the following convenient notation in the context of Bohmian mechanics. Greek letters such as $\phi$ denotes a wavefunction (quantum state), $\phi(q)$ denotes the complex value of the wavefunction $\phi$ at the point $q$ (configuration), $\phi^{*}$ denotes the complex conjugate of the wavefunction, and $\phi^{\dagger}$ denotes the conjugate transpose of the wavefunction with respect to the inner product, so that $\phi^{\dagger}\psi=\langle\phi,\psi\rangle$. Hatted letters such as $\hat{A}$ denote linear operators on the wavefunction space (Hilbert space), and $\hat{A}^{\dagger}$ denotes the adjoint of $\hat{A}$ with respect to the inner product, so that $\langle\psi,\hat{A}^{\dagger}\phi\rangle=\langle\hat{A}\psi,\phi\rangle$. Observables are self-adjoint operators, so that $\hat{A}$ is an observable exactly if $\hat{A}^{\dagger}=\hat{A}$. Bold letters such as $\boldsymbol{x}$ denote three-dimensional column vectors, so that $\boldsymbol{x}=(x_{1},x_{2},x_{3})^{T}$. Arguments of wavefunctions are rank-two tensors, so that for $\phi(q)$ we have \begin{equation} q=(\boldsymbol{x}_{1},\ldots,\boldsymbol{x}_{N})=\left(\begin{array}{ccc} x_{11} & \cdots & x_{1N}\\ x_{21} & \cdots & x_{2N}\\ x_{31} & \cdots & x_{3N} \end{array}\right), \end{equation} which is an element of the tensor space $\mathbbm R^{3\times N}$. The dot product between $3\times N$-tensors $a,b\in\mathbbm R^{3\times N}$ is then defined by \begin{equation} a\cdot b=\sum_{n=1}^{N}\boldsymbol{a}_{n}\cdot\boldsymbol{b}_{n}=\sum_{n=1}^{N}\sum_{k=1}^{3}a_{nk}b_{nk}. \end{equation} Finally, the infinitesimal volume element of the space $\mathbbm R^{3\times N}$ is denoted by $dq=d^{3}x_{1}\cdots d^{3}x_{N}$, so that the integration of a function $f$ over some region $Q=X_{1}\times\cdots\times X_{N}$ with $X_{n}\subset\mathbbm R^{3\times1}$ is written out as \begin{equation} \int_{Q}dq\,f(q)=\int_{X_{1}}d^{3}x_{1}\cdots\int_{X_{N}}d^{3}x_{N}\,f(\boldsymbol{x}_{1},\ldots,\boldsymbol{x}_{N}). \end{equation} The tensor notation of system configurations has the advantage that particle index and the index for the spatial dimension are separate, so that these different concepts are not mingled with each other, which is mathematically convenient and also addresses a criticism put forward by Monton \cite{Monton2002} against wavefunction realism.

\subsection{Foundations of Bohmian mechanics}

A closed system of $N$ spin-free particles is at any time completely described by two physically significant mathematical entities: the configuration $q=(\boldsymbol{x}_{1},\ldots,\boldsymbol{x}_{N})$ of particle positions, and the wavefunction $\Psi$ that guides the particles along their way. Denoting the time parameter by $t\in\mathbbm R$, the temporal evolution of the system is described by the trajectory $\Psi_{t}$ of the wavefunction through the Hilbert space $\mathcal{H}=L^{2}(\mathbbm R^{3\times N})$, and by the trajectory $q_{t}$ of the configuration through the configuration space $\mathcal{Q}=\mathbbm R^{3\times N}$. Both trajectories obey a first-order differential equation: The wavefunction trajectory obeys the \emph{Schr\"odinger equation} \begin{equation} i\frac{d}{dt}\Psi_{t}=\hat{H}\Psi_{t},\label{eq:schroedinger} \end{equation} and the trajectory of the configuration obeys the \emph{guiding equation} \begin{equation} \frac{d}{dt}q_{t}=\frac{j_{t}}{\rho_{t}},\label{eq:guidingeq} \end{equation} where $j_{t}=(\boldsymbol{j}_{1,t},\ldots,\boldsymbol{j}_{N,t})$ is defined by \begin{equation} \boldsymbol{j}_{n,t}=\frac{\hbar}{2m_{n}i}\left(\Psi_{t}^{*}\boldsymbol{\nabla}_{n}\Psi_{t}-\Psi_{t}\boldsymbol{\nabla}_{n}\Psi_{t}^{*}\right),\label{eq:j} \end{equation} and where $\rho_{t}$ is defined by \begin{equation} \rho_{t}=|\Psi_{t}|^{2}.\label{eq:rho} \end{equation} Since the differential equations \eqref{eq:schroedinger} and \eqref{eq:guidingeq} are of first order in time, they have a unique solution for every valid initial condition. More precisely, for every well-behaved initial wavefunction $\Psi_{0}$ at time $t=0$ there is a unique trajectory $\Psi_{t}$ that is formally obtained by applying the unitary time evolution operator $\hat{U}(t)=e^{-i\hat{H}t}$, so that \begin{equation} \Psi_{t}=\hat{U}(t)\Psi_{0}. \end{equation} Similarly, for every initial configuration $q_{0}\in\mathcal{Q}$ for which $\rho_{0}(q_{0})\neq0$, there is a unique trajectory $q_{t}$ that is formally obtained by applying the \emph{trajectory function} $\xi_{t}=\int_{0}^{t}dt'(j_{t'}/\rho_{t'})$, so that \begin{equation} q_{t}=\xi_{t}(q_{0}). \end{equation} Hence, there is a concrete path through space that the particles take, and which is determined by the trajectory function $\xi_{t}$ applied to the initial configuration $q_{0}$ at $t=0$. The path $\boldsymbol{x}_{n}(t)$ of an individual particle $n$ can be extracted from the trajectory function $\xi_{t}=(\boldsymbol{\xi}_{1,t},\ldots,\boldsymbol{\xi}_{N,t})$ by fetching the components corresponding to that particle, so that $\boldsymbol{x}_{n}(t)=\boldsymbol{\xi}_{n,t}(q_{0})$.

These features make Bohmian mechanics a fully deterministic theory. However, there is a probabilistic element introduced into the theory in a manner analog to how probability is introduced into classical mechanics. The observer does not possess the full information about the true initial configuration of the system, but rather he has to retreat to a probability density. According to the \emph{quantum equilibrium hypothesis}, the initial configuration is distributed by the initial density $\rho_{0}=|\Psi_{0}|^{2}$. The Schr\"odinger dynamics allows to conclude that $\rho_{t}=|\Psi_{t}|^{2}$ is the probability density for all times $t$, a feature referred to as \emph{equivariance}. Consequently, the probability to find the system configuration at time $t$ within some region $Q\subset\mathcal{Q}$ is given by \begin{equation} P_{t}(Q)=\int_{Q}dq\,\rho_{t}(q). \end{equation} The functions $\rho_{t}$ and $j_{t}$ can be shown to obey the \emph{continuity equation} \begin{equation} \frac{d}{dt}\rho_{t}+\nabla\cdot j_{t}=0,\label{eq:conteq} \end{equation} so that $j_{t}$ takes the role of a \emph{probability current}. This concludes our brief review of the foundations of Bohmian mechanics. Note that in contrast to standard quantum mechanics, the concept of \emph{measurement} is not an element of the foundations. Rather, measurement is considered as a specially designed but otherwise ordinary physical process that involves both a system of interest and a macroscopic measurement apparatus. Note further that because the guiding equation \eqref{eq:guidingeq} is not local in the position space $\mathbbm R^{3}$, the theory exhibits \emph{quantum nonlocality}. That is, the velocity of each particle generally depends on the instantaneous position of remote particles. This, together with the fact that in contrast to the wavefunction, the positions of the particles are assumed to be unknown to the observer, makes Bohmian mechanics a \emph{nonlocal hidden variables theory}.

\subsection{Measurement in Bohmian mechanics}

In contrast to standard quantum mechanics, there is no separate postulate for measurements in Bohmian mechanics, because a measurement is regarded as a specially designed but otherwise ordinary physical process that involves a short and strong interaction between the system of interest and a macroscopic measurement device involving a large number of particles. The interaction causes the measurement device to change its initial configuration into one out of several macroscopically discernible configurations, each one representing an outcome of the measurement corresponding to a value of the observable to be measured. There is no discontinuous ``collapse of the wavefunction'' but rather a short but continuous unitary evolution of the wavefunction. During that evolution, the system of interest becomes entangled with the measurement device, creating a sum of ``branches'' associated with different potential measurement outcomes. Since the particles of the measurement device are at every instance at a precise position, they occupy exactly one of the branches, and so there is only one unique actual outcome of the measurement. Let us briefly review how this is modeled mathematically.

During a short measurement period $T_{M}$ the system of interest $S$ is coupled to a measurement device $M$ by a strong interaction $\hat{W}_{SM}$, so that the unperturbed Hamiltonian can be neglected, \begin{equation} \hat{H}_{S}+\hat{H}_{M}+\hat{W}_{SM}\approx\hat{W}_{SM}.\label{eq:HW} \end{equation} Before and after the measurement period, the interaction term is zero, so that the system of interest and the measurement device evolve independently. The shortness of the measurement period $T_{M}$ can be more precisely defined by the requirement that the free evolution of the system of interest during a period of length $T_{M}$ can be neglected, that is \begin{equation} e^{-i\hat{H}_{S}T_{M}}\approx\mathbbm1.\label{eq:tmshort} \end{equation} The state of the total system is given by the wavefunction $\Psi=\Psi(x,z)$, where $x=(\boldsymbol{x}_{1},\ldots,\boldsymbol{x}_{N})$ is the variable for the configuration of the system of interest, and $z=(\boldsymbol{z}_{1},\ldots,\boldsymbol{z}_{M})$ is the variable for the configuration of the measurement device. As the system of interest is assumed to be microscopic and the measurement device is assumed to be macroscopic, we have $N\ll M$. The configuration spaces of these systems are $\mathcal{Q}_{S}=\mathbbm R^{3\times N}$ and $\mathcal{Q}_{M}=\mathbbm R^{3\times M}$, respectively, and their volume is measured by the infinitesimal volume elements $dx=d^{3}x_{1}\cdots d^{3}x_{N}$ and $dz=d^{3}z_{1}\cdots d^{3}z_{M}$, respectively. The total configuration space is given by $\mathcal{Q}=\mathcal{Q}_{S}\times\mathcal{Q}_{M}=\mathbbm R^{3\times(N+M)}$, and its elements are the configurations $q=(x,z)$.

Let us give a paradigmatic example with an observable $\hat{A}=\sum_{a}a\,\hat{\varPi}_{a}$ with discrete eigenvalues $a$ and projections $\hat{\varPi}_{a}$ onto the corresponding eigenspaces. With $\eta_{R}$ being the ``ready'' state of the measurement device and $\psi_{a}$ being an eigenstate of $\hat{A}$, the measurement interaction induces for each $a$ the transition \begin{equation} \psi_{a}\otimes\eta_{R}\rightarrow\psi_{a}\otimes\eta_{a},\label{eq:mprocess} \end{equation} where $\eta_{a}$ are macroscopically discernible ``pointer states'', which means that they have almost no spatial overlap, \begin{equation} \eta_{a}(z)\eta_{a'}(z)\approx0\quad\text{for }a\neq a'.\label{eq:nooverlap} \end{equation} Because of \eqref{eq:nooverlap}, and because the pointer states are normalized, there are non-overlapping regions $Z_{a}$ in the pointer space $\mathcal{Q}_{M}$, \begin{align} Z_{a}\cap Z_{a'} & =\emptyset\quad\text{for }a\neq a', \end{align} so that each $\eta_{a}$ has almost all of its support in a corresponding region $Z_{a}$, \begin{equation} \int_{Z_{a}}dz\,|\eta_{a}(z)|^{2}\approx1,\label{eq:quasisupp} \end{equation} and therefore approximately vanishes outside of $Z_{a}$, \begin{equation} \eta_{a}(z\notin Z_{a})\approx0.\label{eq:quasivanish} \end{equation} Let us call $Z_{a}$ an \emph{effective support }of $\eta_{a}$. If the actual configuration of the measurement device comes to lie within the region $Z_{a}$ then this is taken to indicate that ``$a$'' is the outcome of the measurement. Furthermore, the pointer states during free evolution should be \emph{macroscopically stable}, that is, when the pointer configuration comes to lie within a region $Z_{a}$ after measurement, then during a subsequent free evolution it should stay within that region. However, the feature of macroscopic stability is not a necessary requirement.

With these settings, an arbitrary state $\psi$ of the system of interest, together with the initial state $\eta$ of the measurement device, evolves into a sum of ``branches'' corresponding to the measurement device indicating different eigenvalues of $\hat{A}$, \begin{equation} \left(\sum_{a}\psi_{a}\right)\otimes\eta_{R}\rightarrow\sum_{a}\left(\psi_{a}\otimes\eta_{a}\right), \end{equation} where $\psi_{a}=\hat{\varPi}_{a}\psi$ are (unnormalized) eigenstates of $\hat{A}$ corresponding to eigenvalues $a$.

A simple example for such a process \cite{Neumann1932,Bohm1952a,Everett1957} is given by the interaction \begin{equation} \hat{W}_{SM}=-g\hat{A}\hat{p}_{z}, \end{equation} where $g$ is a sufficiently large coupling constant and $\hat{p}_{z}$ is the momentum operator conjugate to the configuration operator $\hat{z}$ of the measurement device. Because of \eqref{eq:HW} the state of the total system after measurement reads \begin{align} \hat{U}(T_{M})\psi\otimes\eta_{R} & =\sum_{a}e^{igaT_{M}\hat{p}_{z}}\hat{\varPi}_{a}\psi\otimes\eta_{R}\\ & =\sum_{a}\psi_{a}\otimes\eta_{a}, \end{align} where the functions \begin{equation} \eta_{a}(z)=\eta(z-gaT_{M}). \end{equation} have their effective support within the respective regions \begin{equation} Z_{a}=\{z\mid z-gaT_{M}\in Z\}, \end{equation} where $Z$ is an effective support of the initial device state $\eta$. Depending on the measurement duration $T_{M}$ and on the separation of the eigenvalues $a$, the coupling constant $g$ must be chosen large enough so that condition \eqref{eq:nooverlap} is met.

Directly after measurement the wavefunction $\Psi'$ of the total system is a sum of branches, \begin{equation} \Psi'=\sum_{a}\Psi'_{a},\label{eq:sumbranch} \end{equation} with each branch \begin{equation} \Psi'_{a}=\hat{\varPi}_{a}\psi\otimes\eta_{a}\label{eq:branch} \end{equation} representing a different potential outcome. However, the configuration of the measurement device can only occupy one of these branches. So what is the probability $P(a)=\text{Prob}(\overline{z}\in Z_{a})$ that the actual configuration $\overline{z}$ of the measurement device comes to lie within the region $Z_{a}$ indicating the outcome ``$a$''? Because of \eqref{eq:nooverlap} we have \begin{equation} |\Psi'(x,z)|^{2}\approx\sum_{a}|\hat{\varPi}_{a}\psi(x)|^{2}|\eta_{a}(z)|^{2}, \end{equation} and thus \begin{align} P(a) & =\int dx\int_{Z_{a}}dz|\Psi'(x,z)|^{2}\\ & \approx\int dx\int_{Z_{a}}dz\sum_{a'}|\hat{\varPi}_{a'}\psi(x)|^{2}|\eta_{a'}(z)|^{2}\\ & \approx\int dx\,|\hat{\varPi}_{a}\psi(x)|^{2}\\ & =\|\hat{\varPi}_{a}\psi\|^{2}, \end{align} where \eqref{eq:nooverlap}, \eqref{eq:quasisupp} and \eqref{eq:quasivanish} have been used. The prediction of Bohmian mechanics thus approximately coincides with the value given by the Born rule of standard quantum mechanics. The quality of the approximation depends on how well the pointer states $\eta_{a}$ are separated in space, which in turn depends on the separation of the eigenvalues $a$, as well as on the strength and duration of the measurement interaction.

Bohmian mechanics is a collapse-free theory, so we may calculate the results of all subsequent measurements by using the time-evolved version of the uncollapsed wavefunction \eqref{eq:sumbranch}. So for $t>t_{M}$, where $t_{M}$ is the time when the measurement is completed, the wavefunction of the total system is given by \begin{equation} \Psi_{t}=\hat{U}(t-t_{M})\Psi'.\label{eq:uncollapsed} \end{equation} The pointer states of a reasonably functioning measurement apparatus should be macroscopically stable, so if the particles only occupy one branch at time $t_{M}$, they will stay on that branch for $t>t_{M}$. However, even if the pointer states were \emph{not} macroscopically stable, the huge number of internal degrees of freedom of a macroscopic measurement device would, as a result of friction and Brownian motion, make a future overlap of the branches so vanishingly small, and the induced quantum interference effects become so highly unlikely that for all practical purposes they can safely be ignored. In other words, the branches \emph{decohere}. Now, since the guiding equation \eqref{eq:guidingeq} is local with respect to the configuration space (although it is \emph{not} local with respect to position space!), the ``empty'' branches of the wavefunction have no influence on the course of the trajectory. Thus, it is a matter of mathematical convenience to ignore the empty branches and only consider, for $t>t_{M}$, an \emph{effectively collapsed} wavefunction given by 
\begin{equation} 
\Psi_{a,t}=\frac{1}{\sqrt{P(a)}}\hat{U}(t-t_{M})\Psi'_{a},\label{eq:effcollaps} 
\end{equation} 
where the normalization has been carried out to preserve the norm of the wavefunction. The ``effective collapse'' in Bohmian mechanics is a mathematical simplification that preserves the predictions for the outcomes of future measurements given the results of past measurements. Physically, the wavefunction remains uncollapsed and it is always possible to use the uncollapsed wavefunction for the prediction of future outcomes.

\bibliographystyle{spphys}

\end{document}